\documentclass[fleqn,twoside]{article}
\usepackage{espcrc2}

\usepackage{graphicx}

\def\simge{
    \mathrel{\rlap{\raise 0.511ex
        \hbox{$>$}}{\lower 0.511ex \hbox{$\sim$}}}}

\title{Pentaquarks: Status and Perspectives for Lattice Calculations}
\author{Shoichi~Sasaki\address[TOKYO]
        {Department of Physics, University of Tokyo, Tokyo 113-0033, Japan} }

\date{}

\begin{document}

\begin{abstract} 
The present status of pentaquark spectroscopy in lattice QCD 
is reviewed. This talk also includes a brief introduction of pentaquark baryons.
\end{abstract}

\maketitle

\section{\label{sec-1}Introduction}

\begin{figure*}[t]
\centering
\includegraphics[width=10.5cm]{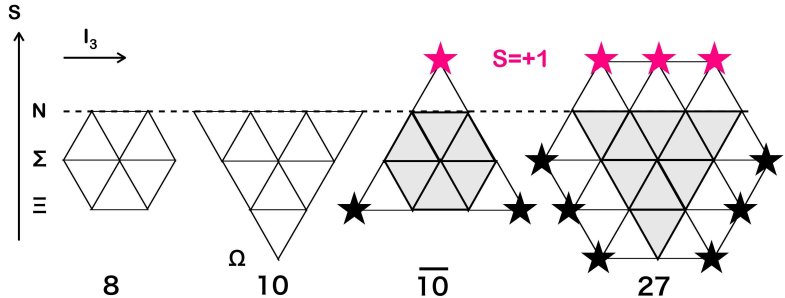}
\vspace{-0.6cm}
\caption{
Weight diagrams for possible pentaquark multiplets. Stars represent the states which have
exotic quantum numbers (electric charge and strangeness).
}
\label{fig:Multplet}
\end{figure*}

The quantum chromo-dynamics (QCD) may not preclude the presence of 
the multi-quark hadrons such as tetraquark ($qq{\bar q}{\bar q}$), 
pentaquark ($qqqq{\bar q}$), dibaryon ($qqqqqq$) and so on, 
because of the color confinement.
Especially, we are let to be interested in {\it exotic 
multi-quark hadrons}, which should have {\it exotic} quantum numbers.  
Now, let us address ourself to the pentaquark state. Consider the $SU(3)$ flavor case.
Pentaquark states should form six different multiplets:
%
%
\[
3_{f} \otimes 3_{f} \otimes 3_{f} \otimes 3_{f} \otimes \overline{3}_{f}
\]
\[
\;\;\;\;\;\;\;\;\;\;\;=
1_{3} \oplus 8_{8} \oplus 10_{4} \oplus \overline{10}_{2} \oplus 27_{3}
\oplus 35_{1}
\]
where subscripts in the right hand side denote the number of degeneracy in each multiplet.
The first three multiplets are common in the case of usual baryons.
However, the last three multiplets; antidecuplet, 27-plet and 35-plet, 
are distinct irreducible representations since those multiplets 
have an apparent exotic quantum-number such as {\it strangeness +1}.
Needless to say, $S=+1$ baryon can not be accommodated by usual baryons.
Other possible exotic quantum-numbers are represented as stars in Fig.~\ref{fig:Multplet}.
Of course,  one cannot predict 
which multiplet is preferred for the possible $S=+1$ pentaquark baryon,
within the group theoretical argument.
The Skyrme model~\cite{Skyrmion} and 
the chiral soliton model~\cite{Diakonov:1997mm}, 
however, predict that the lowest $S=+1$ state appears 
uniquely in the antidecuplet and its spin and parity is 
spin-half and positive parity. 

Recently, the LEPS collaboration at Spring-8 has observed a very sharp peak resonance 
in the $K^{-}$ missing-mass spectrum of the 
$\gamma n \rightarrow nK^{+}K^{-}$ reaction on $^{12}C$~\cite{Nakano:bh}. 
The observed resonance should have strangeness +1. Thus, $\Theta^{+}(1540)$ 
cannot be a three quark state and should be an exotic baryon 
state with the minimal quark content $uudd \bar s$.
The peak position is located at 1540 MeV with a 
very narrow width. Those are quite consistent with the chiral-soliton model's 
prediction~\cite{Diakonov:1997mm}. 
This discovery is subsequently confirmed by 
other experiments~\cite{{EXPs},{Barth:2003es}}\footnote{ 
It should be noted, however, that the experimental evidence
for the $\Theta^{+}(1540)$ is not very solid yet since there
are a similar number of negative results to be reported~\cite{Hicks:2004vd}.}.
Experimentally, spin, parity and isospin are not 
determined yet. Non-existence of a narrow resonance in $pK^+$ 
channel indicates that possibility of $I=1$ has been already 
ruled out~\cite{{Nakano:bh},{Barth:2003es}}

%
%

Many theoretical studies of pentaquarks are also triggered by the discovery 
of the $\Theta^{+}(1540)$.
I introduce the most reputed model proposed by Jaffe and Wilczek~\cite{Jaffe:2003sg}. 
In the naive quark models, the low-lying pentaquark state should have spin-1/2 and 
negative parity. However, in this case, the pentaquark baryon just falls apart into
$KN$ in a S-wave. It is difficult to explain its very narrow width.
Jaffe and Wilczek propose a simple idea to flip the parity of  the low-lying 
pentaquark. Suppose there is the strong diquark correlation.
The spin-0, color triplet and flavor triplet diquark would be favored within the
simple one gluon exchange. The pentaquark can be composed of two 
identical bosons (diquarks)
and one antiquark. However, the anti-symmetrization in terms of color, requires
relative odd number's angular-momentum between the pairs of identical bosons. 
Otherwise, the wave function of the pentaquark state should be vanished.
Resulting parity of the low-lying pentaquark is same 
as the chiral soliton model~\cite{Jaffe:2003sg}.
In addition, the $S=+1$ baryon is uniquely assigned to antidecuplet in this description.

Consequently, correlated quark models, {\it i.e.} the diquark model, 
may accommodate the positive-parity 
isosinglet pentaquark $\Theta(uudd{\bar s})$ as same as the chiral soliton model.
However, there are essential differences between the diquark model and the
chiral soliton model. In correlated quark models, one cannot fully discriminate 
the other multiplet such as octet. If apart from the $SU(3)$ flavor limit, 
the mixing between octet and antidecuplet should be taken into account.
Jaffe and Wilczek advocate the ideal mixing case, which is favored in the vector meson spectrum~\cite{Jaffe:2003sg}.
The ideal mixing provides the different prediction for the mass of possible exotic 
$\Xi^{--}_{3/2}$ state, which should have isospin-3/2.
Furthermore, the Roper resonance, $N(1440)$, can be accommodated 
in the diquark model owing to the ideal mixing~\cite{Jaffe:2003sg}.

A candidate of the isospin-3/2 pentaquark state
has been reported by the NA49 
collaboration~\cite{Alt:2003vb}. 
They observed the sharp peak resonance, which has exotic quantum numbers, in $\Xi \pi$ 
missing mass analysis. The observed mass is located somewhat between predictions 
made by 
the chiral soliton model and the diquark model.
In this experiment, the isospin partner, $\Xi^{0}_{3/2}$ is also observed.
It should be pointed out that this discovery is not confirmed 
yet by other experiments~\cite{Hicks:2004vd}.

I shortly mention about charm or bottom analog of the $\Theta$ state.
If anti-strange quark is replaced by a heavier antiquark, what is going to be happened. 
Several models predict that charm or bottom analog of the pentaquark is expected to 
be a bound state~\cite{{Jaffe:2003sg},{anti-charm}}. 
Recently, the discovery of the 
exotic anti-charmed baryon ($uudd{\bar c}$) has been announced by 
the H1 collaboration~\cite{Aktas:2004qf}. However, the observed state is higher than 
the $DN$ threshold and also the $D^{*}N$ threshold against 
several model predictions~\cite{{Jaffe:2003sg},{anti-charm}}.
Again, this discovery has not been confirmed yet by other experiments.

\section{\label{sec-2}Lattice pentaquark spectroscopy}

 If the pentaquark baryons really exist, such states must emerge directly from first principles, QCD. Of course, what we should do is to confirm the presence of the
pentaquarks by lattice QCD. Experimentally, it is rather difficult to 
determine the parity of the $\Theta^{+}(1540)$. Thus, lattice QCD has a 
chance to answer the undetermined quantum numbers before experimental efforts. 
Lattice QCD has also a feasibility to predict the masses for undiscovered pentaquark
baryons. I stress that there is substantial progress in lattice study of excited 
baryons recently~\cite{Sasaki:2003xc}.
Especially, the negative parity nucleon $N^*(1535)$, which lies close
to the $\Theta^{+}(1540)$, has become an established state 
in quenched lattice QCD~\cite{{Sasaki:2003xc},{Sasaki:2001nf}}.
Here I report that quenched lattice QCD is capable of studying the $\Theta^{+}(1540)$ as well.

Indeed, it is not so easy to deal with the $qqqq{\bar q}$ state 
rather than usual baryons ($qqq$) and mesons ($q \bar q$) in lattice QCD.
The $qqqq{\bar q}$ state can be decomposed into a pair of color singlet states
as $qqq$ and $q \bar q$, in other words, can decay into 
two hadron states
even in the quenched approximation.
For instance, one can start a study with a simple minded local operator for the $\Theta^{+}(1540)$,
which is constructed from the product of a neutron operator and a $K^{+}$ 
operator such as ${\cal O}= \varepsilon_{abc}(d^{T}_{a}C\gamma_5 u_{b})d_{c}({\bar s}_{e}\gamma_5 u_{e})$.
The two-point correlation function composed of this operator,
in general, couples not only to the $\Theta$ state (single hadron)
but also to the two hadron states such as 
an interacting $KN$ system~\cite{{Luscher:1986pf},{Fukugita:1994ve}}.
Even worse, when the mass of the $qqqq{\bar q}$ state is
higher than the threshold of the hadronic two-body system, the two-point function
should be dominated by the two hadron states.
Thus, a specific operator with as
little overlap with the hadronic two-body states 
as possible is desired in order to identify the 
signal of the pentaquark state in lattice QCD.
Once one can identify the pentaquark signal in lattice QCD, 
to determine the parity of the $\Theta^{+}(1540)$ is the most 
challenging issue at present. Thus, it is necessary to project out
the parity eigenstate from given lattice data precisely.
In the following subsections, I discuss three 
related issues

%
%
\begin{figure}[t]
\centering
\includegraphics[width=7.0cm]{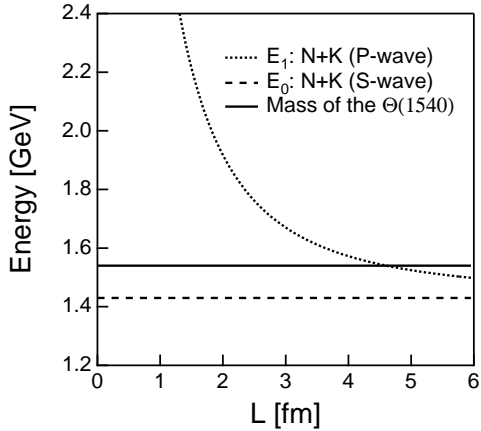}
\vspace{-0.6cm}
\caption{
The S-wave and P-wave $KN$ threshold energies 
on a lattice of spatial extent $L$.
In the case of $L \le 4.6$ fm, the mass of the $\Theta^{+}(1540)$ is lower 
than the P-wave $KN$ threshold.}
\label{fig:SpectraFlow}
\end{figure}  

\subsection{\label{sec-2.1} Estimation of the $KN$ threshold}

The experimentally observed $\Theta^{+}$(1540) state is clearly a resonance state. 
However, its mass is near the $KN$ threshold. We could manage to calculate the pentaquark
as a bound state {\it if its parity were positive}. 

I recall that all momenta are quantized as $\vec{p}L=2\pi \vec{l}$
(${\vec l} \in {\bf Z}^3$) on lattice in finite volume (the spacial extent $L$) 
with the periodic boundary condition (PBC)
.\footnote{Of course, this quantization
 condition may change to $\vec{p}L+2\delta(p)=2\pi \vec{l}$
 with a scattering phase shift $\delta(p)$ by an interaction between two hadrons. 
However, the $KN$ channel has very weak interactions as is well known experimentally.
One may omit this shift for a crude estimation. } 
Thus, the spectrum of energies of two hadron states such as
$KN$ states with zero total momentum should be discrete and 
these energies are approximately equal to values, which are evaluated
in the noninteracting case:
\begin{equation}
E^{KN}_{n}=\sqrt{M^2_{N}+{p}^2_{n}}+\sqrt{M^2_{K}+{p}^2_{n}}
\label{eq:energies}
\end{equation}
where $p_{n}=\sqrt{n}\cdot {2\pi/L}$ and $n\in {\bf Z}$.
The positive parity $\Theta$ state decay into $KN$ in a P-wave 
where the $KN$ system should have a nonzero relative momentum.
The P-wave $KN$ threshold is simply estimated at  an energy level
$E_{1}$, which is evaluated with the smallest nonzero momentum 
$p_1=2\pi/L$ in Eq. (\ref{eq:energies}).
 This energy level $E_{1}$ can be lifted by decreasing spatial extent as 
 depicted in Fig.~\ref{fig:SpectraFlow} while the lowest energy level $E_{0}$, which
 corresponds to the S-wave $KN$ threshold, remains unchanged.
 The level crossing between $E_{1}$ and the $\Theta$ mass takes place around
 4.6 fm in this crude estimation. It implies that {\it the positive parity $\Theta$ state 
 may become a bound state in the typical lattice size of currently available lattice 
 calculations, i.e. $L \approx 2-3$ fm}. 

\subsection{\label{sec-2.2} Choice of operators}

For the case of the negative parity $\Theta$ state, the presence of the 
$KN$ scattering state must complicates the study of pentaquarks in lattice QCD. 
One should choose an optimal operator, which couples weakly to the 
$KN$ scattering state, in order to access the pentaquark state above the
(S-wave) $KN$ threshold.

\begin{figure}[t]
\centering
\includegraphics[width=7.0cm]{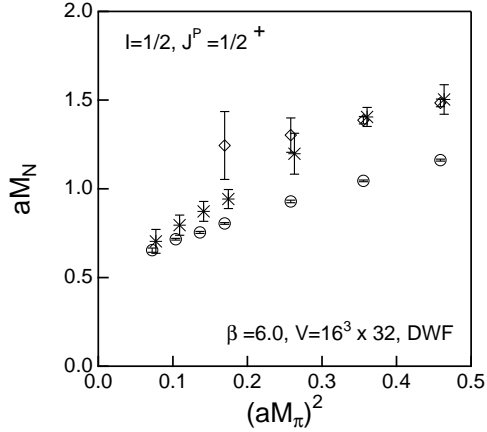}
\vspace{-0.6cm}
\caption{Comparison of the fitted mass from 
$\langle{\cal O}^{N}_{1}\overline{\cal O}{}_1^{N}\rangle$ ($\circ$),  
$\langle{\cal O}^{N}_{2}\overline{\cal O}{}_2^{N}\rangle$ ($\diamond$)
and the cross correlation $\langle{\cal O}^{N}_{1}\overline{\cal O}{}_2^{N}
+{\cal O}^{N}_{2}\overline{\cal O}{}_1^{N}\rangle$ ($\ast$) \cite{Sasaki:2001nf}.}
\label{fig:IntOpt}
\end{figure}

For this direction, I would like to recall that the less known observation in
the spectroscopy of the nucleon. 
There are two possible interpolating operators
for the $I=1/2$ and $J^{P}=1/2^{+}$ state;
${\cal O}^{N}_{1}=\varepsilon_{abc}[u^{T}_a C\gamma_5 d_b] u_c$ and 
${\cal O}^{N}_{2}=\varepsilon_{abc}[u^{T}_a C d_b] \gamma_5 u_c$, 
even if one restricts operators to contain no derivatives and to belong to the $(\frac{1}{2},0)\oplus(0,\frac{1}{2})$
chiral multiplet under $SU(2)_{L}\otimes SU(2)_{R}$~\cite{Sasaki:2001nf}.
Of course, two operators have the same quantum number of 
the nucleon. The first operator ${\cal O}^{N}_{1}$ is utilized conventionally 
in lattice QCD since the second operator 
${\cal O}^{N}_{2}$ vanishes in the non-relativistic limit. It implies that the second operator is 
expected to have small overlap with the nucleon $|\langle 0|{\cal O}^{N}_{2}
|{\rm Nucl}\rangle|  \approx 0$. 
Indeed, the mass extracted from the correlator constructed 
by the second operator ${\cal O}^{N}_{2}$ exhibit the different mass from 
the nucleon mass~\cite{{Sasaki:2001nf},{OptDep}} as shown in Fig.~\ref{fig:IntOpt}.
The operator dependence on a overlap with desired state is evident,
at least, in the heavy quark regime, while the cross correlation 
suggests that the small overlap with the nucleon might be 
no longer robust in the light quark regime where is far from the 
non-relativistic description~\cite{Sasaki:2001nf}. 

%

\subsection{\label{sec-2.3}Parity projection}
The intrinsic parity of the {\it  local} baryon operator can be defined 
by the parity transformation of internal quark fields as
%
%
\begin{equation}
{\cal P}{\cal O}^{(\eta)}({\vec x},t){\cal P}^{\dagger}
=\eta \gamma_{4}{\cal O}^{(\eta)}(-{\vec x},t)\;\;,
\end{equation}
where $\eta =\pm 1$.
However, due to the relation ${\cal O}^{(+)}=\gamma_5{\cal O}^{(-)}$
for the {\it local} baryon operator, 
the resulting two-point correlation functions are also related with 
each other as $\langle {\cal O}^{(+)}(x)\overline{\cal O}{}^{(+)}(0)\rangle
=-\gamma_5\langle {\cal O}^{(-)}(x)\overline{\cal O}{}^{(-)}(0)\rangle\gamma_5
$.
This means that the two-point correlation function composed of 
the {\it local} baryon operator can couple to
both positive- and negative-parity states. However, I note that
{\it anti-particle contributions of opposite parity states is propagating forward 
in time}. 
Thus, the $+/-$ parity eigenstate in the forward propagating contributions
is obtained by choosing the appropriate parity 
projection $(1\pm \eta \gamma_4)/2$, which is 
given in reference to the intrinsic parity of operators, $\eta$.
Details of the parity projection are described in Ref.~\cite{Sasaki:2001nf}.

\section{\label{sec-3}First exploratory studies}

\subsection{\label{sec-3.1}Local pentaquark operators}

As I remarked previously, an optimal operator, which couples 
weakly to the $KN$ scattering state, would be required to explore 
the pentaquark baryons in lattice QCD.
For this purpose, two types of local pentaquark operator for isosinglet state 
are proposed in the first two studies.
One is {\it a color variant} of the simple product of nucleon and kaon operators, 
%
%
\begin{equation}
{\cal O}_{I=0}^{(-)}=\varepsilon_{abc}[u_a^{T}C\gamma_5 d_b]\{u_e({\bar s}_e \gamma_5 d_c)
- (u \leftrightarrow d) \},
\end{equation}
which is proposed by Csikor {\it et al.}~\cite{Csikor:2003ng} 

The other is proposed by Sasaki~\cite{Sasaki:2003gi} as in
a rather exotic description guided by the diquark-diquark-antiquark structure:
%
%
\begin{equation}
{\cal O}_{I=0}^{(\eta)}=
\varepsilon_{abc}\varepsilon_{aef}\varepsilon_{bgh}[u_{e}^{T}C\Gamma_{1} d_{f}]
[u_g^{T}C\Gamma_2 d_h]C{\bar s}_c^T
\end{equation}
where $\Gamma_{1, 2}= 1, \gamma_5, \gamma_5\gamma_{\mu}$ (but $\Gamma_1 \neq \Gamma_2$)
and the superscript ``$\eta$" stands for the intrinsic parity of the operator.
There are three kinds of diquark-diquark-antiquark operator in this description, which are 
useful for the extended study with the $3\times 3$ correlation matrix analysis.
More details of construction of the diquark-diquark-antiquark operator are described in
Ref.~\cite{Sasaki:2003gi}.

In an exploratory study, one may assume that those interpolating operators have smaller overlap with 
the $KN$-scattering state than the simple product of nucleon and kaon operators, at least, in the heavy quark 
regime. Because, in the non-relativistic limit, all of them give rise to the different wave function 
from the $KN$ two-hadron system.

%
%
\begin{figure}[t]
\centering
\includegraphics[width=7.0cm]{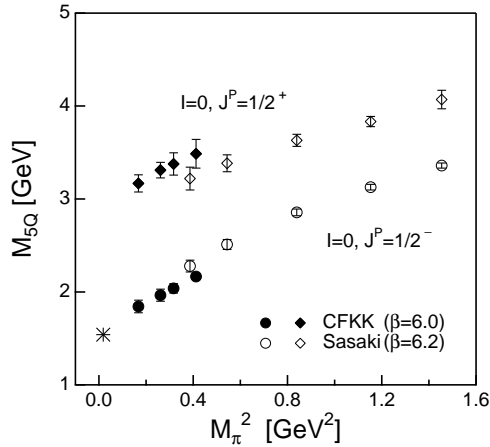}
\vspace{-0.6cm}
\caption{
Masses of the isosinglet
$S=+1$ baryons with both positive- and negative-parity
as functions of pion mass squared~\cite{{Csikor:2003ng},{Sasaki:2003gi}}.
The experimental value for $\Theta^{+}(1540)$ is marked with a star.
}
\label{fig:CFKKSasaki}
\end{figure}  

\subsection{\label{sec-3.2}Results}

The first two lattice studies were performed
with the Wilson gauge action and the Wilson fermion action
at the almost same box size $L\simeq 2.0 - 2.2$ fm.
The lattice spacing Csikor {\it et al.} use is rather coarse than
that of Sasaki, but their calculation was employed with relatively
lighter pion masses ($M_{\pi}\sim 0.4-0.6$ GeV). The main difference
between two studies is the choice of pentaquark operators.

After some initial confusion about the parity assignment\footnote{
See a footnote in Ref.~\cite{Csikor:2003ng}. },
both calculations agreed that the lowest state of the isosinglet
$S=+1$ baryons has the negative parity as shown in Fig.~\ref{fig:CFKKSasaki}.
The main results from the first two exploratory studies can be summarized 
as follows.
\begin{itemize}
\item The current lattice simulations seem to give no indication of a pentaquark
in the positive parity channel to be identified with the $\Theta^{+}(1540)$.

\item The negative parity channel can easily accommodate a pentaquark
with a mass close to the experimental value.
\end{itemize}
Therefore, both authors conclude that the exploratory lattice study
favors spin-parity  $(1/2)^-$ and isospin 0 for the $\Theta^{+}(1540)$.
In Ref. \cite{Sasaki:2003gi}, the anti-charmed analog of the $\Theta$ state 
was also explored. It is found that the $\Theta_c(uudd{\bar c})$ lies 
much higher than the $DN$ threshold, in contrast to 
several model predictions~\cite{{Jaffe:2003sg},{anti-charm}}.
More detailed lattice study would be desirable to clarify the significance of
those observations.

%
%
\begin{table*}[ht]
\caption{Summary of the present status of lattice pentaquark spectroscopy}
\label{table:1}
\newcommand{\m}{\hphantom{$-$}}
\newcommand{\cc}[1]{\multicolumn{1}{c}{#1}}
\renewcommand{\arraystretch}{1.2} 
\begin{tabular}{@{}lcccc}
\hline
author(s) and reference           & signal & parity of pentaquark   & kind of operator \\
\hline
\hline
Csikor {\it et al.}~\cite{Csikor:2003ng} & Yes & negative & color variant of $KN$ \\
Sasaki~\cite{Sasaki:2003gi}                 & Yes & negative & diquark-diquark-antiquark \\
Mathur {\it et al.}~\cite{Mathur:2004jr} & No   & N/A & simple $KN$ \\
Chiu-Hsieh~\cite{Chiu:2004gg}        & Yes & positive & diquark-diquark-antiquark \\
Ishii {\it et al.}~\cite{Ishii:2004qe}         & No   & N/A & diquark-diquark-antiquark \\
Alexandrou {\it et al.}~\cite{Alexandrou:2004ws} 
                                                                    & Yes & negative & diquark-diquark-antiquark \\
MIT group~\cite{MIT}                               & Yes & negative & diquark-diquark-antiquark \\
YITP group~\cite{YITP}            & Yes & negative & simple $KN$ and color variant of $KN$\\
\hline
\end{tabular}
\end{table*}

\section{\label{sec-4} Subsequent lattice studies}

There are four subsequent lattice studies of pentaquark 
spectroscopy to be found in the 
literature~\cite{{Mathur:2004jr},{Chiu:2004gg},{Ishii:2004qe},{Alexandrou:2004ws}}.
Other two preliminary results had been also 
reported at some conferences~\cite{{MIT},{YITP}}.
I give a short review of those results as follows.

Recently, Kentucky group performed their simulations near the physical pion 
mass region with overlap fermions~\cite{Mathur:2004jr}.  
However, they choose {\it the simple minded operator as the product of nucleon 
and kaon operators} to explore the pentaquark baryons.
Instead, the sequential constrained fitting method is applied in their analysis
to disentangle the pentaquark signal from towers of $KN$ scattering state.
They also check carefully the volume dependence of the spectral weight. 
They claim as follows: Only towers of $KN$ scattering state are seen 
in the negative parity channel.
They confirmed that ground state in either parity channels has a characteristic
volume dependence on the spectral weight, which should have $1/L^3$ dependence for
two particles. They also confirm that the ghost contribution from $KN \eta'$ state appears 
from pion mass less than around 0.3 GeV in the positive parity channel.
Their final conclusion is that there is no sign of pentaquark signal in either parity 
channels in their calculation. However, it seems that their results are consistent with the
experimental fact that the $\Theta^{+}(1540)$ state has not yet been found
in the $KN$ scattering data as an elastic resonance~\cite{Arndt:2003xz}.

Another negative results against the first two studies are reported 
by TIT group~\cite{Ishii:2004qe}.
First, they trace calculations of Ref.~\cite{Sasaki:2003gi} on an anisotropic 
lattice with the $O(a)$ improved Wilson fermion action. 
They confirm that the lowest energy state appears in the negative
parity channel. In their study, a new method is proposed  to lift up the S-wave $KN$ 
threshold by imposing a hybrid boundary condition (HBC) in the spatial direction
and it is also applied in their calculation. They found that the plateau of the lowest energy 
state is shifted in the effective mass plot by changing the spatial boundary condition 
from PBC to HBC as is expected in the case of the two hadron states. However, their calculations 
are employed with smeared sources, which are apparently optimized for the ground 
state of usual hadrons, {\it i.e.} nucleon and kaon.
This optimization may enhance the signal of the $KN$ scattering states
rather than that of the pentaquark state. Indeed, the resulting effective mass plots for the pentaquark
do not show any effectiveness of their smearing in the earlier Euclidean time region, 
in contrast to the case of usual baryons. However, of course, their method, {\it i.e.} 
the hybrid boundary method, is quite promising to distinguish between the pentaquark
state and the $KN$ scattering state. 

There are several positive results for the first two studies.
Cyprus group performed careful studies concerning the volume dependence of
the spectral weight and then found that their observed state seems to be a single
hadron state, {\it i.e.} the pentaquark state~\cite{Alexandrou:2004ws}.
MIT group formulates eight possible local operators based on
the diquark-diquark-antiquark structure to perform the $4\times 4$ correlation matrix 
analysis in both isosinglet and isotriplet channels as a extension of Sasaki's proposal~\cite{MIT}.
Their preliminary results are currently obtained from double exponential fits on
data for each single operator.
They confirmed results of the first two studies at their exploratory stage.
YITP group has started the $2 \times 2$ correlation matrix analysis with
the simple minded operator as the product of nucleon and kaon operators 
and its color variant~\cite{YITP}.
Their preliminary results support that there is an indication of the presence of the pentaquark 
state near the lowest $KN$ scattering state in the negative parity channel. 

Finally, I comment on Chiu-Hsieh's results. 
In the first version of Ref.~\cite{Chiu:2004gg}, they 
initially misunderstood the parity assignment. Then, their results
are completely opposite to ones of any other studies where the lowest
energy state appears in the negative parity channel. In the second version,
they corrected this apparent error and reanalyzed their data. However,
their final conclusion remains unchanged. They insist that 
the spin-parity of the $\Theta^{+}(1540)$ state is most likely $(1/2)^{+}$.
This conclusion seems to attribute to their crude chiral extrapolation.

\section{\label{sec-5} Summary and Outlook}

Table~\ref{table:1} represents a summary of the present status for
each lattice calculation.
The first conclusion of the first two studies as summarized in Sec.~\ref{sec-3.2}
is confirmed by subsequent lattice studies. The currently important issue is 
whether or not to establish the presence of the $\Theta^{+}(1540)$ in the negative 
parity channel. It is necessary for this 
to disentangle the pentaquark signal from the $KN$ scattering states completely .
The correlation matrix analysis is strongly required to separate the $KN$ scattering state and 
isolate the pentaquark state. We also should check the volume dependence of the
spectral weight. Probably, the hybrid boundary condition is helpful to identify the
pentaquark state as a single hadron state. Needless to say, we ought to try non-local 
 types of pentaquark operator in order to verify whether there is no indication of
 the $\Theta^{+}(1540)$ in the positive parity channel.

Finally I stress that all present results should be regarded as 
exploratory. Indeed, much detail studies are in progress in each group.
Thus, I had better to conclude that the following questions still 
remain open:
%
%
%
\begin{itemize}
\item Does the spectrum of QCD possess the $\Theta^{+}(1540$)?
\item What is spin and parity of the $\Theta^{+}(1540$)?
\item Are there other pentaquark baryons, {\it e.g.} the charm (bottom) pentaquark or 
the spin-orbit partner of the $\Theta^{+}(1540)$?
\end{itemize}
%
There are many exiting issues to be explored.

\section*{Acknowledgments}

This talk has benefited from conversation and correspondence with
T.-W. Chiu, Z. Fodor, S. Katz, F.-X. Lee, K.-F. Liu, N. Mathur and J. Negele.
The author is supported by JSPS Grant-in-Aid for Encouragement of 
Young Scientists (No. 15740137).


\end{document}